# SPACE-CHARGE EFFECTS IN FIELD EMISSION: ONE DIMENSIONAL THEORY

A. Rokhlenko [a], K. L. Jensen [b], J. L. Lebowitz [a,c]

[a] Dept. of Mathematics, Rutgers University, Piscataway, New Jersey 08854-8019

[b] Code 6843, ESTD, Naval Research Laboratory, Washington, DC 20375-5347

[a] Dept. of Physics, Rutgers University, Piscataway, New Jersey 08854-8019

The current associated with field emission is greatly dependent on the electric field at the emitting electrode. This field is a combination of the electric field in vacuum and the space charge created by the current. The latter becomes more important as the current density increases. Here, a study is performed using a modified classical 1D Child-Langmuir description that allows for exact solutions in order to characterize the contributions due to space charge. Methods to connect the 1D approach to an array of periodic 3D structures are considered.

---

[†]  79.70.+q   Field emission, ionization, evaporation, and desorption
  85.45.Db   Field emitters and arrays, cold electron emitters
  85.45.Bz   Vacuum microelectronic device characterization, design, and modeling
  52.59.Sa   Space-charge-dominated beams

# INTRODUCTION

Space charge, or the effects of Coulomb interactions between emitted electrons, and its impact on the generation and transport of electron beams are important in many practical applications, particularly when beam brightness and modulation characteristics are critical to the operation of the device. Consequently, both space charge and emittance dominated beams have received intense study, particularly as they relate to thermal and photoemission sources for amplifiers and particle accelerators [1-6]. In contrast, field emission relies on quantum mechanical tunneling through barriers thinned through the application of fields on the order of 10 GV/m at the emission site. As the highest macroscopic fields that can be generated are smaller (100 MV/m for rf photoinjectors injectors, 1-50 MV/m for vacuum electronic devices, *etc*.) field enhancement must be utilized by making the emission sites have sharp features [7, 8]. While field emitters that resemble pyramids, ellipsoids, or whiskers [9] in microfabricated arrays [10] boast current densities at the emission sites many orders of magnitude larger than for thermal and photoemission sources, the emission area is generally on the order of 100 $nm^2$ per site. The high curvature of the site geometry makes modeling of field emitters in modern PIC codes an issue: simple models of space charge or emittance from these sources are unable to fully exploit the one-dimensional techniques often used in the analysis of other electron sources, or treat many separate emitters acting in concert. Nevertheless, there have been several theoretical efforts [11-16]: in particular, the treatment of the 1D field emission / space charge problem (Ref. [17] and references therein), the usage of Particle-in-Cell codes to treat field emission (Ref. [18] and references therein) and the discussion of numerical issues involved in modeling space-charge-limited flow (even in the 1D case using PIC codes can lead unusual behavior as a consequence of cell-size related problems) [19], that are pertinent.

The present study is the first of a two part effort to describe the coupling of 1D and 3D approaches to field emission in a manner that allows for the effects of space charge on the operation of microfabricated structures to be addressed. Here, a methodology for the analysis of space charge forces on the emission process is described, and a manner in which 1D methods may be brought to bear on arrays of emitters operating together is presented (the second, and separate, study concerns the impact of space charge on individual 3D structures, and builds on methods introduced in the analysis of dark current [20] and emittance [21]. The objective is to provide a framework to investigate field emitters without intensive numerical efforts in a manner amenable to PIC codes for when space charge is an issue and the cathode area is but a small region of the simulation. Applications that rely on field emission will benefit, and such applications include (but are not limited to): electron beam lithography [22, 23] and transmission electron microscopes [24]; spacecraft propulsion [25, 26]; mm-wave Vacuum Electronic amplifiers and THz devices [27, 28]; and particle accelerators and Free Electron Lasers (FEL's) [29-31].

## THE 1D MODEL

Poisson's equation in one dimension can be written

$$\frac{d^2}{dx^2}V(x) = \frac{q^2}{\varepsilon_o}\rho \quad (1)$$

where $q$ is the elementary charge, $\rho$ is a number density, and $V(x)$ is potential energy. The current density $J$ is given by

$$J = qv\rho = q\rho\left(\frac{2V}{m}\right)^{1/2} \quad (2)$$

where (as in Table 1) $v$ is the electron velocity and $m$ is the electron mass. The second equality comes from the assumption that the velocity is zero at the cathode where $V = 0$. Introduce the dimensionless quantities $\varphi$, $y$, and $j$ such that

$$\begin{aligned} V &= \varphi V_o \\ J &= jJ_o \\ x &= yL \end{aligned} \quad (3)$$

where $V_o$ is the anode potential, $J_o$ is a characteristic current, and $L$ is the distance between the anode and the cathode surface. Eq. (1) becomes (compare to the treatments of Refs. [32, 33])

$$\frac{d^2}{dy^2}\varphi(y) = \frac{j}{\sqrt{\varphi}} \quad (4)$$

where $J_o$ is defined by

$$J_o \equiv \frac{9}{4} J_{cl}(V_o, L) = \sqrt{\frac{2}{m}} \left( \frac{\varepsilon_0 V_o^{3/2}}{qL^2} \right) \qquad (5)$$

and $J_{cl}$ is the familiar Child-Langmuir maximum current that can be drawn across an anode-cathode gap with no initial velocity of the electrons and zero field at the cathode [3]. Solving Eq. (4) with appropriate boundary conditions is straightforward and has been done in the literature (see Ref. [17] and references therein), but the solution is synopsized in the present notation with an eye to eliminate problems with the appearance of non-physical results. Representing the product of charge $q$ and field at the cathode by $F = fV_o/L$, where $f$ is dimensionless, solutions to Eq. (4) are then given by

$$\frac{d\varphi}{dy} = \left( 4j\sqrt{\varphi} + f^2 \right)^{1/2} \qquad (6)$$

A second integration yields (compare Eq. 3 of Ref. [11])

$$\left( 4j\sqrt{\varphi} + f^2 \right)^{1/2} \left( 2j\sqrt{\varphi} - f^2 \right) + f^3 = 6j^2 y \qquad (7)$$

Invoking the boundary conditions that at $y = 1$, $\varphi = 1$, then gives

$$\left( 4j + f^2 \right)^{1/2} \left( 2j - f^2 \right) + f^3 = 6j^2 \qquad (8)$$

Eq. (8) is an exact equation for the 1D diode and does not have unphysical solutions: manipulations with it should be done with care in order not to introduce them. Eq. (8) can be written as

$$j = \frac{1}{9} \left[ 2 + (2 - 3f)\sqrt{1 + 3f} \right] \qquad (9)$$

which provides a universal relation between the current and the field at the cathode independent of the relation between the field at the cathode and the emitted current (see also Forbes [17]). It encapsulates the two most familiar limits: in the limit that the current vanishes, then $f = 1$, or the field at the cathode $F = V_o/L$; conversely, if $f = 0$, then $j = 4/9$ (the Child-Langmuir, or CL, current density). Observe that the field $f$ at the cathode cannot exceed 1 (its vacuum value) when space charge is absent and there is no screening.

We shall consider particular solutions depending on the behavior of $j(f)$, namely, a linear and quadratic dependence, which can be handled analytically, and a Fowler-Nordheim linear field model [34]. Using the image charge, or Schottky barrier lowering, factor [35] is reserved for the 3D case, and shall be considered separately [36]. Before considering these cases, we note that Eq. (8) can be manipulated and re-expressed in a convenient form without square roots as

$$3f^2(1-f) - j(4-9j) = 0 \tag{10}$$

which bears a relationship to Eq. (13) of Forbes [17] (note however that Eq. (10) allows unphysical solutions as the second root of the resulting quadratic equation, e.g., $f = 1$ and $j = 4/9$, satisfies Eq. (10) but is not a solution of Eq. (8)).

**Linear Current-Field Relationship**

For $j = af$, that is, the current is linear in the cathode surface field, then according to Eq. (10) the field satisfies

$$0 = 3f^2(1-f) - af(4-9af) \tag{11}$$

The physically relevant solution is given by

$$f = \frac{1}{2}\left\{1 + 3a^2 + (1-3a)\sqrt{1+\frac{2}{3}a+a^2}\right\} \tag{12}$$

which yields $f = 1$ when $a = 0$ (and thus $j = 0$). The case $a = 1/3$ corresponds to $f = 2/3$ and $j = 2/9$ (half of the space charge limit). The behavior of Eq. (12) is shown in Figure 1 in a form that shows the asymptotic convergence of $f$ to the CL limit: the y-axis is the ratio of $j = af$ with the space charge limit $j_{CL} = 4/9$. The increasing influence of space charge can be seen in the small $a$ limit, whereas the approach to space charge limited current can be seen in the large $a$ limit: these limits are

$$\begin{aligned} j(a \ll 1) &\approx \frac{a}{1+\frac{4}{3}a+\frac{5}{9}a^2} \\ j(a \gg 1) &\approx \frac{4}{9+\frac{3}{a^2}} \end{aligned} \tag{13}$$

The expansions are reasonably good: at $a = 1$ Eq. (13) suggests that $j \approx 0.346$ and $0.333$ for the small and large expansions, respectively, as compared to the exact value $j = 0.367$,

**Quadratic Current-Field**

When $j = af^2$, then Eq. (10) becomes

$$0 = f^2\left[3(1-f) - a(4-9af^2)\right] \tag{14}$$

The physical solution of Eq. (14) is

$$f = \frac{1}{6a^2}\left[1-(1-2a)\sqrt{4a+1}\right] \quad (15)$$

The behavior of the current $j = af^2$ found from Eq. (15) is shown in Figure 2 in a form that exhibits its asymptotic convergence to the CL limit as done in the linear case. As before, the influence of space charge in the small and large $a$ limits are

$$j(a \ll 1) \approx \frac{a}{1+\frac{8}{3}a-\frac{2}{3}a^2}$$

$$j(a \gg 1) \approx \frac{4}{9+\frac{27}{4a}} \quad (16)$$

where Eq. (16) suggests that at $a = 1$, $j \approx 0.333$ and $0.254$ for the small and large expansions, respectively, as compared to the exact value $j = 0.291$. Also, observe that $f$ is approximately unity for small values of $a$: specifically, to leading order in $a$, then $f(a \ll 1) \approx 1-(4a/3)$ for both the linear and quadratic cases.

The question of what values of the parameters determine when the current becomes space-charge dominated is very important in practice. For the 1D geometry and the simple emission models considered, the answers appear in Figures 1 and 2: the current is half of the CL limit (*i.e.* $j= 2/9$) when $a = 1/3$ and $1/2$ for the linear and quadratic dependences respectively. These quantities can be easily related to physical units by using Eq. (3).

**Fowler-Nordheim Current-Field Relation**

For the triangular barrier, the Fowler Nordheim current-density versus field relationship $J_{FN}(F)$ (*i.e.,* Eq. 8 of Ref. [35]) can be expressed as

$$j = \frac{J_{FN}(F)}{J_o} \equiv af^2 \exp\left(-\frac{b}{f}\right) \quad (17)$$

where

$$a = \frac{1}{\pi(\mu+\Phi)}\sqrt{\frac{\mu R_\infty V_o}{\Phi}}$$

$$b = \frac{4}{3}\left(\frac{L}{a_o}\right)\left(\frac{\Phi}{V_o}\right)\sqrt{\frac{\Phi}{R_\infty}} \quad (18)$$

and terms are as defined in Table 1. The work function $\Phi$ is, in the absence of the image charge, equivalent to the height of the triangular barrier above the Fermi energy $\mu$ for a metal. The form of Eq. (18) makes the dimensionless nature of $a$ and $b$ transparent. Their dependence on the anode potential $V_o$ may give pause, but recall that the current density depends on the field strength at the cathode (*i.e.*, $f$) rather than the anode potential, and that $j$ is scaled by the CL result. Eq. (8) can be written as

$$\left(4j+f^{2}\right)^{1/2} = \frac{6j^{2}-f^{3}}{2j-f^{2}} \tag{19}$$

The denominator vanishes for $j = f^2/2$. For the left hand side is finite, the numerator must vanish, or $f^3 = (3/2)f^4$. Thus, the degenerate root in the FN case is given by $f = 2/3$ and $j = 2/9$ (half of the CL limit). In MKSA units, this will occur when:

$$V_{trans} = \frac{2\Phi\left(\dfrac{L}{a_o}\right)\sqrt{\dfrac{\Phi}{R_\infty}}}{\ln\left(\dfrac{2}{\pi(\mu+\Phi)}\sqrt{\dfrac{\mu R_\infty V_{trans}}{\Phi}}\right)} \tag{20}$$

which can be solved iteratively. For example, for $L = 1$ µm, $\mu = 7$ eV and $\Phi = 4$ eV, then $V_{trans}$ converges to 21.953 keV after 8 iterations for a starting guess of 10 keV. $V_{trans}$ therefore serves as a transition point between when one branch of the numerical solution of $j_{FN}(f)$ is taken ($V < V_{trans}$) versus the other ($V > V_{trans}$): this affects, for example, the maximum and minimum $V$ considered in a bisection method to find the current density, as in the first branch $V_{trans}$ is the larger value, whereas in the second branch, it is the smaller. For an $L$ of 1 micron, the impact of space charge on the current density versus anode potential is shown in Figure 3 for metal-like parameters. Field emission results are more conventionally shown on an FN plot of $1/V$ versus $\ln\left(J_{FN}(V)/V^2\right)$, and so the data of Figure 3 is recast in that form in Figure 4 with $V$ in keV and $J$ in A/cm$^2$. The impact of lowering the work function to 2.0 eV and 0.5 eV is shown in Figures 5 and 6, respectively. The departure from linearity on such plots indicates the impact of space-charge forces.

**Current Regimes: Limiting Cases**

Small Current

Small values of $a$ in the linear, quadratic, and more general cases such as Fowler-Nordheim correspond to small $j$, and therefore, Eqs. (13) and (16) for $j$ for more general relations can be obtained from an expansion of Eq. (9) for small $(1 - f)$: to second order

$$j(f \approx 1) \approx \frac{3}{4}(1-f) - \frac{15}{64}(1-f)^2 \qquad (21)$$

In conjunction with a Taylor expansion of $j(f)$ for $f$ near unity given by

$$j(f) = j(1) - (1-f)j'(1) \qquad (22)$$

where prime indicates derivative with respect to argument, then to leading order,

$$j(f) \approx \frac{j(1)}{1 + \frac{4}{3}j'(1)} \equiv \frac{j(1)}{1+\varepsilon} \qquad (23)$$

which is equivalent (to order $a$) to Eqs. (13) and (16) with regards to the denominator. For the Fowler-Nordeim $j(f)$ then Eqs. (17) and (23) imply

$$\varepsilon_{FN} \approx \frac{8}{3}ae^{-b} \qquad (24)$$

and arises from increasing space charge when the cathode efficiency (larger $a$ and smaller $b$) is larger.

The correction factor $\varepsilon_{FN}$ that arises from space charge is not difficult to examine experimentally because strong field emission is not required. Importantly, if the anode potential $V_o$ and the anode-cathode separation $L$ are changed such that their ratio $F = V_o/L$ remains constant, then by virtue of the definition of $\varepsilon_{FN}$ and $j$, it is seen

$$\varepsilon_{FN} = \frac{1}{3}(b+2)\frac{J_{FN}(F)}{J_{CL}\left(V_o, L = \frac{V_o}{F}\right)} \qquad (25)$$

and so because $J_{CL}(V_o, F/V_o)$ scales as $V_o^{-1/2}$, it is seen that $\varepsilon_{FN}$ scales as $V_o^{1/2}$.

Large Current

When the current density is large and $j(f)$ of a more complex form (*e.g.*, Eq. (17)), then numerical methods are required to solve Eq. (8). However, if emission is so large that $f$ is small, then from Eq. (9) it follows

$$j = j_o - f^2 \frac{\left(2+\sqrt{3f+1}\right)}{\left(1+\sqrt{3f+1}\right)^2} \approx j_o - \frac{3}{4}f^2 \qquad (26)$$

where (from Eq. (5)) $j_o = 4/9$. That is, when $f > 1/4$, then $j$ is greater than 90% of the CL limit without much regard to the details of the dependence of $j$ on $f$.

For small $f$ and constant $V_o/L$, then the variation in $j$ with $a$ for the linear ($l$), quadratic ($q$), and FN relationships behave as

$$\begin{aligned} \partial_a j_l &= f \\ \partial_a j_q &= f^2 \\ \partial_a j_{FN} &= f^2 e^{-b/f} \end{aligned} \qquad (27)$$

We note that for small $f$ there is a decrease in the magnitude of these derivatives which manifests iself in a slower approach to the CL limit (see Figures 1 and 2).

**Comparison to PIC Simulations**

Particle-in-Cell, or PIC, codes are widely used methods to analyze the impact of space charge forces on beam transport. Electrons can be launched into the anode-cathode gap region, and their presence introduces fields that impede subsequent emission, especially if the emission is dependent on the field at the cathode surface. Although time dependent phenomena, particularly in the form of oscillations in the current density, are to be expected, such variations damp out and a steady state equilibrium can be approached. PIC simulations addressing field emission have been performed by Feng and Verboncouer [18]: they examined the approach of field emission to space charge limited flow in a 1D simulation. Here, the findings of Feng and Verboncoeur are compared to predictions based on solutions to Eq. (9).

There are differences between the emission model used by Feng and Verboncoeur and the triangular barrier model of the original Fowler-Nordheim equation: an image-charge corrected form of the Fowler-Nordheim equation is employed by the PIC simulations in which the barrier is lowered by a Schottky factor by $\Delta\Phi = \sqrt{\alpha_{fs}\hbar cF}$. Therefore, the height of the barrier above the Fermi level $\mu$ is given by

$$\phi = \Phi - \sqrt{\alpha_{fs}\hbar cF} \equiv (1-y)\Phi \tag{28}$$

which defines the term $y$. As discussed by Murphy and Good [8], the primary impact of the Schottky factor is to alter the $a$ and $b$ factors in Eq. (17) by appending $t(y)^{-2}$ to the former and $v(y)$ to the latter (and will be treated in greater detail in the second study [36]): the intention here is to compare the 1D theory of Eq. (10) to the PIC simulations of Feng and Verboncoeur, and so their approximations shall be used temporarily. However, as discussed by Forbes [37], $F_{max} = \Phi^2/\alpha_{fs}\hbar c$ corresponds to that field which lowers the emission barrier to the Fermi level (*i.e.*, $\phi = 0$): thus, in the usage of the FN equation with image charge correction, fields larger than $F_{max}$ are disallowed by the approximations from whence the FN equation is derived [38]. If the largest field considered is 10 GV/m, then the work function must be larger than 3.8 eV. As a result, only the findings summarized in Figure 11 of Ref. [18] for which the work function is 4 eV and the anode-cathode separation is 1 μm are considered.

As shown elsewhere [39, 40], the impact due to changes of a tunneling barrier shape on the exponential term in the FN equation are captured in the factor $v(y)$ (for the image charge potential) or analogs to it. That is, the image charge barrier FN equation can be made to resemble the triangular barrier FN equation *if the "effective" work function $\Phi_{eff}$ depends on the field F*. From Eq. (18), it is seen that the exponential term of the triangular barrier FN equation scales as $b \propto \Phi^{3/2}$. Equating the exponential arguments of the triangular barrier FN equation with the image charge form, it is seen

$$\Phi_{eff}^{3/2} = \Phi^{3/2} v\left(\frac{1}{\Phi}\sqrt{\alpha_{fs}\hbar cF}\right) \tag{29}$$

The coefficient of the exponential terms in both the triangular and image charge FN equations have different dependences on the work function parameter. In keeping with the "effective" work function concept, a factor of $C$ is appended to the triangular FN equation. A naive equivalence between the triangular and effective equations then suggests

$$C \approx \frac{1}{4\Phi t(y)^2} \left( \frac{\Phi_{eff}}{\mu} \right)^{1/2} \left( \mu + \Phi_{eff} \right) \tag{30}$$

which, for copper-like parameters and fields of 3 GV/m (the image charge barrier more closely resembles a triangular barrier for low field) is $C \approx 0.42$. Because the equivalence is not exact (an "effective" work function only roughly relates an image charge barrier to a triangular barrier) it can only be concluded that $C$ is on the order of 1/2.

Solutions of Eq. (10) using Eq. (17) for $j(f)$, and Eq. (18) for $a$ and $b$ for which $\Phi$ is replaced by $\Phi_{eff}$ as per Eq. (29) were calculated. The final $j$ was then scaled by the factor $C = 1/2$. Observe that scaling by $C$ does *not* make the scaled $j$ (the solution of Eq. (9)) the same as the space charge limited FN current found by Feng and Verboncouer: the scaling is *only* to show that the 1D triangular barrier FN equation used in Eq. 9 anticipates the image charge FN equation in that their dependence on $F$ is qualitatively comparable. Using the approximation Feng and Verboncouer use of $v_{FV}(y) \approx 1 - y^{1.69}$, the results of the comparison are shown in Figure 7: the combination of an "effective" work function and a scaling parameter anticipates the PIC simulations remarkably well, demonstrating that the 1D solution using the triangular barrier FN equation is in fact a good predictor (albeit not a replacement) of the behavior to be found using the image charge FN equation in in PIC simulations to within a scale factor $C$ for the space charge limited current.

The triangular barrier FN equation was used so that small work function potentials could be considered without running afoul of the Schottky factor lowering the emission barrier to below the Fermi level. Not only for small barriers, but for barriers in general, the $j$ of Eq. (9) can be suitably modified to accommodate situations where temperature and/or field make the emitted current density complex and possibly temperature dependent (*e.g.,* the General Thermal Field Equation given by Eq. (36) of Ref. [41]).

**Connection of the 1D Model to an Array**

The 1D model regards emission of uniform sheets of charge from a surface [42]. Field emission arrays, in contrast, emit from a lattice of emitter sites. As is well-known in electrostatics (and explicitly utilized by the Point Charge Model [20, 21]), conducting surfaces can be replaced by equipotential surfaces at the same potential, and the converse holds as well: the equipotential surface signifying "anode" in the point charge model becomes the "cathode" of the 1D region, as schematically illustrated in Figure 8. How far from the plane of charges the "anode" in the point charge model must be before the discrete nature of the PCM is sufficiently smoothed to approximate the planar "cathode" in the 1D approach is now investigated.

Consider a sheet of point charges spaced on a square grid for which the tip-to-tip distance is $a_{tt}$ (alternately, the pitch of the array) and the magnitude of each charge is proportional to $\lambda$: by superposition, the sheets due to the other $\lambda$'s can be considered separately and combined afterwards. Express all lengths in units of $a_{tt}$ and to take the cathode surface to be at $z = 0$ and the anode to be at $z = N/2$ and far away ($N \gg 1$). For the anode potential to be constant and uniform, image charges of equal and opposite sign to the cathode point charges are placed at $z = N$. The potential everywhere between $0 < z < N/2$ is then given by

$$V(x,y,z) = V_0 + \frac{\lambda}{4\pi\varepsilon_0 a_{tt}} \sum_{j=-\infty}^{\infty} \sum_{k=-\infty}^{\infty} \varphi_{jk}(x,y,z)$$

$$\varphi_{jk}(x,y,z) = \frac{1}{R_c} - \frac{1}{R_a} = \frac{R_a^2 - R_c^2}{R_a R_c (R_a + R_c)} = \frac{N(N-2z)}{R_c R_a (R_c + R_a)} \quad (31)$$

$$R_c = \left[(j-x)^2 + (k-y)^2 + z^2\right]^{1/2}$$

$$R_a = \left[(j-x)^2 + (k-y)^2 + (N-z)^2\right]^{1/2}$$

where the ordering of the arguments in $\varphi$ reflect that the potential satisfies $V(x+n, y+m, z) = V(x,y,z)$ for integer $n$ and $m$, and therefore $x$ and $y$ may be assumed less than 1/2. For computational purposes, the last form of $\varphi_{i,j}$ in Eq. (31) is preferred. Likewise, the $z$ component of the field is given by

$$\hat{z} \cdot \mathbf{F}(x,y,z) = \frac{\lambda}{4\pi\varepsilon_0 a_{tt}^2} \sum_{j=-\infty}^{\infty} \sum_{k=-\infty}^{\infty} \varphi'_{jk}(x,y,z)$$

$$\varphi'_{jk}(x,y,z) \equiv \frac{z}{R_c^3} + \frac{N-z}{R_a^3} \quad (32)$$

The other components are analogous, but the $z$ component is dominant. How large must $z$ be before the point particle nature of the charges is obscured sufficiently to approximate the 1D framework is the question.

Consider a finite circular array of point charges of diameter $2M$ such that Eq. (31) is the $M \to \infty$ limit. It is expedient to numerically analyze finite $M$ and generalize to the $M \to \infty$ case. For $z$ sufficiently large that the series can be well approximated by an integral, it follows

$$\hat{z} \cdot \mathbf{F} = \lim_{M \to \infty} F_M$$

$$\frac{F_M}{F_o} = \frac{1}{4\pi} \sum_{j=-M}^{M} \sum_{k=-\sqrt{M^2-j^2}}^{\sqrt{M^2-j^2}} \varphi'_{j,k}(z)$$

$$\approx \frac{1}{2} \int_0^M \left[ \frac{zr}{(r^2+z^2)^{3/2}} + \frac{(N-z)r}{(r^2+(N-z)^2)^{3/2}} \right] dr \qquad (33)$$

$$= 1 - \frac{z}{2\sqrt{M^2+z^2}} - \frac{(N-z)}{2\sqrt{M^2+(N-z)^2}}$$

where $F_o = \lambda / \varepsilon_0 a_{tt}^2$. It is seen that the second two terms in the third line of Eq. (33) are equivalent to

$$\frac{z}{2\sqrt{M^2+z^2}} + \frac{(N-z)}{2\sqrt{M^2+(N-z)^2}} = \frac{1}{2} \int_M^\infty r \, dr \, \varphi'_{jk}(z) \bigg|_{j^2+k^2=r^2} \qquad (34)$$

For $M \gg N > z$, then to leading order the $z$ component of field is $F = F_o[1-(N/2M)]$. Clearly, as $M \to \infty$, $F$ approaches the parallel plate solution $F_o$. The finite $M$ form of Eq. (32) can be evaluated numerically, and how large $z$ should be can be inferred from when Eq. (33) approaches its $1 - (N/2M)$ limit. Consider the variation of $F_M/F_o$ given by Eq. (33) along the line $x = y$, for which variations are at their maximum, for representative values of $(N,M)$ of (5,10) and (10,40) in Figure 9. For $z > 1$, the undulations in $F_z$ are largely absent. This can also be seen in the behavior of the potential itself in Figure 10 for $(N,M)$ values of (4,8) at $z = 0.4$ (A) and $z = 1.6$ (B): the color scales differ as the potential increases with $z$, but it is clear that the collection of point-like potentials for $z < 1$ merges into a more uniform, or sheet-like, behavior for $z > 1$. The small values of $N$ and $M$ are to facilitate visual inspection, but field emitter arrays are characterized by far larger $M$ than considered herein, for which $F_z$ is more nearly constant within the disk defining the active array area. Consequently, $z$ values of order unity are the appropriate locations for which to transition from a unit cell representation (discussed separately [36]) to the 1D approach, with the "anode" of former being the "cathode" plane of the latter. Issues remain about the optimal choice of the field of that boundary, and how to account for its small but nevertheless present variation with $x$ and $y$, but those questions shall be taken up in a separate study. For present purposes, the 1D $f$ is related to the 3D $F_o$ via $f \propto F_o$ and $j \propto I_{tip} / a_{tt}^2$.

A demonstration of how rapidly the potential (or electric field) directly above a point charge converges with the potential (or electric field) above the midpoint of four point charges provides the final indication of how rapidly the ripples in Figures 9 and 10 decline. As seen in Eq. (31) for the potential and Eq. (32) for the field, it will amount to finding numerically how rapidly the potential off-axis and on-axis approach the same value. From Eq. (31), it is seen that such a question can be most easily answered by ignoring Vo and evaluating

$$R_\varphi(z) = \frac{\sum_{j^2+k^2<M^2} \varphi_{jk}(0,0,z) + \delta_M}{\sum_{j^2+k^2<M^2} \varphi_{jk}\left(\frac{1}{2},\frac{1}{2},z\right) + \delta_M} - 1$$

$$R_F(z) = \frac{\sum_{j^2+k^2<M^2} \varphi'_{jk}(0,0,z) + \delta'_M}{\sum_{j^2+k^2<M^2} \varphi'_{jk}\left(\frac{1}{2},\frac{1}{2},z\right) + \delta'_M} - 1 \qquad (35)$$

Eq. (35) will approach 0 as $z$ increases. It is seen that adding $4\pi\varepsilon_0 a_{tt} V_0 / \lambda$ to both the numerator and denominator in the fractional part (where the $M \to \infty$ limits give terms leading to the potential and field from $R_\varphi$ and $R_F$, respectively) will only cause $R_\varphi$ and $R_F$ to decrease more rapidly. Therefore, Eq. (35) is a good metric to consider how the ripples fade. The terms $\delta_M$ and $\delta'_M$, evaluated analogously to Eq. (34), appear in both the numerator and denominator, as for large $M$, whether $x = y = 0$ or $1/2$ is of small importance. We find

$$\delta_M = \frac{1}{2}\left\{\left[M^2 + (N-z)^2\right]^{1/2} - \left[M^2 + z^2\right]^{1/2}\right\}$$

$$\delta'_M = \frac{1}{2}\left\{\frac{z}{\left[M^2+(N-z)^2\right]^{3/2}} + \frac{N-z}{\left[M^2+z^2\right]^{3/2}}\right\} \qquad (36)$$

The behavior of $R_\varphi(z)$ and $R_F(z)$ are shown in Figure 11 as a function of $z$ where $M$ is taken to be 20 (for comparison, $M = 10$ is also shown, and $N = 10$ for both): it therefore is a measure of the amplitude of the ripples of Figure 9 and demonstrates the rapidity with which those ripples vanish. The field ripples vanish less rapidly than the potential ripples because $\varphi'$ is more sensitive to variations than $\varphi$.

With a minor modification, it is possible to consider a cathode composed of sharper tips in a background field $F_o$. Let the cathode plane be at $z = 0$ as before, but now consider two oppositely charged particles symmetrically placed about the $z$ plane for each emitter which we shall call the "dipole" model (it anticipates the "dipole" model that shall be examined at length in the single-tip space charge investigation considered separately), in contrast to Eq. (35) that uses only one charge to represent the tip ("monopole" model). Therefore, the potential of the dipole in a background field is given by

$$V_{dipl}(x,y,z) = F_o z + \frac{\lambda}{4\pi\varepsilon_0 a_{tt}} \sum_{j=-\infty}^{\infty} \sum_{k=-\infty}^{\infty} \varphi_{jk}^{dipl}(x,y,z)$$

$$\varphi_{jk}^{dipl}(x,y,z) = \frac{1}{R_+} - \frac{1}{R_-} \quad (37)$$

$$R_{\pm} = \left[(j-x)^2 + (k-y)^2 + (z\pm d)^2\right]^{1/2}$$

In other words, the anode at $z = N/2$ in the monopole model has been replaced by a background field $F_o$. As done with $Vo$ in the monopole model, the background field $F_o$ may be neglected in the dipole analog of Eq. (35), as the ratios minus 1 will be smaller with $F_o$ than without it, and so the calculation without $F_o$ serves as an upper bound. It is seen that the dipole model formulae for $R_\varphi$ and $R_F$ are obtained by the replacements $N \to 2d$ and $z \to z+d$ in the monopole formulae for $R$ and $\delta$, but the interpretations are different: $d$ in the dipole model is smaller than the tip-to-tip spacing, whereas $N$ in the monopole model is much greater than the tip-to-tip spacing. Performing the analogous calculation for the dipole model, the results are shown in Figure 12 using $M = 20$.

**Gated Structures**

A final complication is the possible presence of a gate near the field emitter, which for FEA's serves to modulate the array: "emission-gated" current is required by a variety of technological applications from displays to microwave amplifiers [9] and FEL's [31]. The simplest model of the gate (a metal plane collinear with the emitter apexes from which disks centered about the emitter apex are excised, as shown in the expanded diagram in Figure 8 of the gated cell) is to add to the charges that represent the emitter a charged ring to represent the gate (the so-called Saturn Model [43]). The potential is then, in spherical coordinates,

$$\varphi_{saturn}(r,\theta) = \frac{q}{4\pi\varepsilon_0}\left\{\frac{q_{tip}}{r} - \frac{q_{ring}}{r}\sum_{l=0}^{\infty}(-1)^l\frac{(2l)!}{2^{2l}(l!)^2}P_{2l}(\cos\theta)\right\} \qquad (38)$$

where $P_l(\cos\theta)$ is a Legendre polynomial, and where the magnitudes of the charges $q_{tip}$ and $q_{ring}$ are comparable. Therefore $\varphi$ decreases faster than in the ungated case when the ring is absent, to leading order as $(q_{tip} - q_{ring})/r$, or, if $q_{tip} = q_{ring}$, as $2q_{tip}a_g^2/r^3$ for small $\theta$. Therefore, an array of Saturn rings plus charges should coalesce more rapidly to the 1D representation than the ordered array of bare point charges considered previously.

For both gated and ungated geometries, an approach that may allow for the estimation of the impact of space charge on field emission from arrays in a manner amenable to particle-in-cell (PIC) beam simulation codes is suggested. PIC is used to model the injection and acceleration of charge bunches especially when space charge complicates transport and causes emittance growth [19, 44-48]. Indeed, the needs of PIC codes suggest using the PCM approach to develop the "cathode" boundary of the PIC simulation. In so framing the problem, the space charge limits considered in the 1D section of this work have direct bearing to more comprehensive and time dependent PIC simulations, there by allowing field emission sources to be treated by methods that in the past have been profitably used primarily on planar cathode structures.

## UNIVERSAL FEATURES OF FIELD EMISSION IN 1D

In spite of its simplicity, the 1D Child-Langmuir (CL) model for a cathode of unlimited emissivity plays a very important role in the design and analysis of various devices [1] as well as for developing effective computational schemes for more realistic geometries. The situation is more complicated in the case of field emission because the current density depends on the field at the cathode surface, but surprisingly, Eq. (9) allows one to find universal features of emission which are independent of specific cathode properties (as also done by Forbes [17], who found a simplified version in his Eq. (15), though Eq. (9) here does not introduce extraneous roots). As noticed by Forbes, and shown by Eq. (9), $j$ is a function of $f$ only, and in particular, how close the current produced by field emission approaches the CL limit is determined solely by the field at the cathode surface (in units of the vacuum field). To show this, a plot of $9j/4$, with $j$ given by Eq. (9), is compared to the example lines $j_L = f$, $j_Q = f^2$, and $j_{FN} = f^2 \exp(-1/f)$ is shown in Figure 13. The intersection points of the former with the three latter lines gives the value of $f$ that solves Eq. (9) for the particular current-field relation (linear, quadratic, or Fowler Nordheim, respectively).

As demonstrated in Figure 12, $9j/4 > 0.9$ as soon as $f < 0.27$ (alternately, $F < 0.27 V_o / L$), and even when $f = 1/2$, the current is not far from the CL limit ($9j/4 \approx 0.697$). the conclusion that a field with a magnitude of a quarter of the vacuum field $V_o/L$ is sufficient to make emission almost equal to the CL limit can be observed in experiments. When $9j/4$ approaches unity and $f$ approaches zero, then the following relationship is obtained from Eq. (9) for $F \ll V_o / L$:

$$F^2 \approx \left(\frac{8mq^2}{9\varepsilon_0^2}\right)^{1/2} V_o^{1/2} \left[J_{CL}(V_o, L) - J(F)\right] \tag{39}$$

where we have returned to dimensioned units and use has been made of Eq. (5). Thus, for a given CL limit, $F$ scales as $V_o^{1/4}$.

## CONCLUSION

The relation of space charge to field emission is an important problem because of the strong variation of the emitted current with the field that exists at the emission site. Consequently, charge between anode and the emitter (or emitter-gate) boundary bears a complex relation to the voltages and separation distances defining the diode region. Three studies of space charge and its impact on field emission are therefore indicated. First, a 1D analysis of the basic Poisson relation in a diode region is required. Second, a representation of a 3D field emitter structure adaptable to the 1D theory is needed. Third, and finally, a method to approximate the boundary needed in the 1D problem by considering the behavior of a 3D periodic array is needed. In the present study, the first and third issues have been addressed, and the second shall be reported separately. Although the present analysis is steady state, its results are commensurate with the asymptotic behavior of the PIC simulations of Feng and Verboncouer [18]. The transition from 3D to 1D for the case in which field emission arises from a surface that has irregularities or is purposely modified in such a way that the average spacing between the irregularities (spikes) is $H$, than at a height $H$ above the surface of the cathode toward the anode, the electric field is indistinguishable from the 1D field, which is of consequence when cell-size constraints that are already important in PIC codes (e.g., Ref. [19]) are an issue.

**Acknowledgements**

The work of AR and JLL was supported in part by AFOSR Grant AF-FA 9550-07; they thank Dr. R. Barker for useful discussions. The work of KLJ was supported provided by the *Naval Research Laboratory* and the *Joint Technology Office*; he thanks K. Nguyen (*Beam Wave Research*) and J. Petillo (*SAIC*) for useful discussions.

Table 1: Symbols and Parameters

| Symbol | Definition | Value | Unit |
|---|---|---|---|
| *Fundamental Constants and Parameters* | | | |
| $m$ | Electron Mass | 510999 | eV/c² |
| $c$ | Speed Of Light | 2997.92 | nm/fs |
| $\hbar$ | Planck's Constant | 0.658212 | eV-fs |
| $q$ | Unit Charge | 1 | q |
| $R_\infty$ | Rydberg Constant | 13.6057 | eV |
| $\alpha_{fs}$ | Fine Structure Constant | 1/137.036 | - |
| $a_o$ | Bohr Radius $\hbar/\alpha_{fs} mc$ | 0.0529177 | nm |
| $\varepsilon_0$ | Permitivity Of Free Space | 5.52635x10⁻² | q²/eV-nm |
| $Q$ | $\alpha_{fs}\hbar c/4$ | 0.359991 | eV-nm |
| $F$ | Field At Cathode | - | eV/nm |
| $V$ | Potential | - | eV |
| *Copper-like Parameters* | | | |
| $\mu$ | Chemical Potential | 7 | eV |
| $k_F$ | $(2m\mu)^{1/2}/\hbar$ | 13.5546 | 1/nm |
| $v_F$ | Fermi velocity | 1.56919 | nm/fs |
| $\Phi$ | Work Function | 4.5 | eV |
| *Fowler Nordheim Field Emission Parameters* | | | |
| $B$ | $J_{FN}$ Parameter Eq. 20 | 65.2073 | eV/nm |
| $\kappa$ | $J_{FN}$ Parameter Eq. 20 | 0.772808 | - |
| $v(y)$ | Elliptical integral function | $1-y^2\left(1-\frac{1}{3}\ln(y)\right)$ | - |
| $t(y)$ | Elliptical integral function | $\approx t(y_o) = 1.061$ | - |
| $y_o$ | $e^{-1/2}$ | 0.606531 | - |

# FIGURE CAPTIONS

**Figure 1**

Behavior of $f$ as given by Eq. (12) for the linear current-field relationship (the grey dashed line represents the CL limit).

**Figure 2**

Behavior of $f$ as given in Eq. (15) for the quadratic current-field relationship (the grey dashed line represents the CL limit).

**Figure 3**

Onset of space charge effects for a high (4.0 eV) work function for the 1D field emission current density assuming a 1 micron anode to cathode separation, metal-like parameters, and a triangular barrier (no Schottky barrier lowering - Eq. (17)). The dashed line represents the Fowler Nordheim Equation with $F = V_o/L$; the solid thick line therefore corresponds to $F = fV_o/L$ and therefore includes the effects of space charge.

**Figure 4**

Same as Figure 3, but for the data represented on a traditional Fowler-Nordheim plot.

**Figure 5**

Same as Figure 4, but for a middle (2.0 eV) work function. The thin line labeled "CL" is the Child-Langmuir limit (see Eq. (5)).

**Figure 6**

Same as Figure 4, but for a small (0.5 eV) work function and small (0.5 eV) chemical potential.

**Figure 7**

Comparison of the space charge limited current as a function of applied field (ratio of anode voltage to anode-cathode separation) data of Figure 11a of Ref. [18] ("Feng &

Verboncouer") to Eq. (10) using the effective work function $\Phi_{eff}$ of Eq. (29) without ("unscaled JFN") and with ("scaled JFN") $C = 1/2$: see discussion following Eq. (30).

**Figure 8**

Schemmatic of an array of emitters in the ungated (left) and gated (right) configuration. The 1D approach applies for large $z$ (top) and the unit cell approach for small $z$ (bottom). An enlargement of the gated emitter is shown on the bottom right, and its Saturn Model representation on the bottom left, where the point and ring charges are in the gate plane.

**Figure 9**

The $z$ component of the gradient of the potential along the diagonal defined by $x = y$ for a circular emitters arranged on a square grid with $(N,M) = (5,10)$ for the left (A), and (10,40) for the right (B) for various values of $z$ (red = 0.4, blue = 0.8, black =1.6). The length is scaled by half of the diagonal of the total square region (that is, $\sqrt{2}M$). For infinite uniform charged parallel plates, the field between them would be constant (independent of $z$) and equal to $F_o$.

**Figure 10**

Same as Figure 9, but showing the potential itself over the array for $(N,M) = (4,8)$ and $z = 0.4$ (A) and 1.6 (B). The solid color in the center of B shows that as $z$ increased beyond unity, the potential near the center is more uniform.

**Figure 11**

Magnitude of the ripple amplitude functions $R_\varphi$ and $R_F$ as a function of distance $z$ from the cathode surface, where $z$ is in units of the tip-to-tip spacing of the array, for the "monopole" model of Eq. (35).

**Figure 12**

Magnitude of the ripple amplitude functions $R_\varphi$ and $R_F$ as a function of distance $z - d$, where $z$ is in units of the tip-to-tip spacing of the array, for the "dipole" model of Eq. (37).

**Figure 13**

The ratio of $9j/4$, where $j$ is the ratio of current density with the CL limit, as a function of $f$ (ratio of surface field $F$ to vacuum field $V_o/L$) for various current-field relationships: $L$ = linear; $Q$ = quadratic, and $FN$ = Fowler Nordheim, for the cases $a = b = 1$.

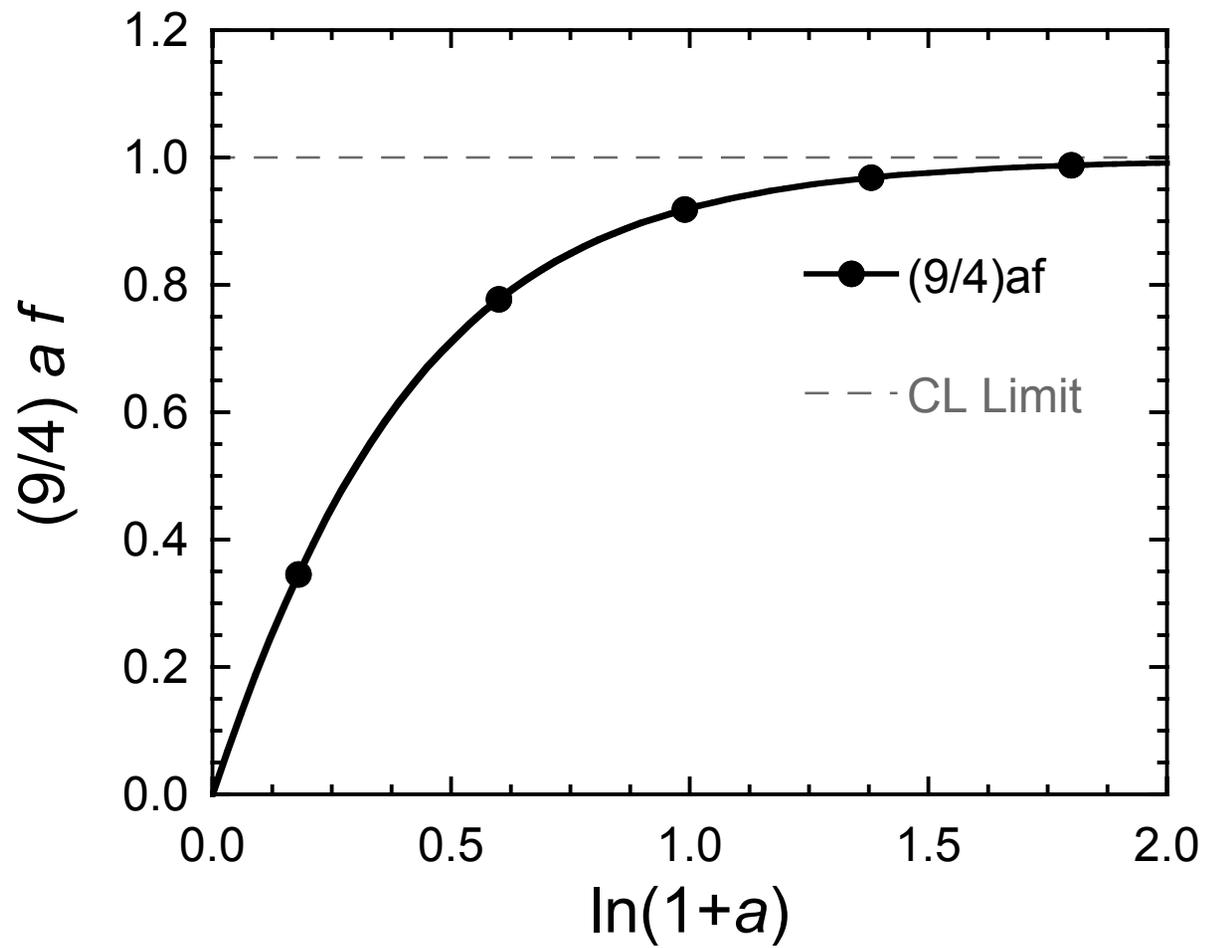

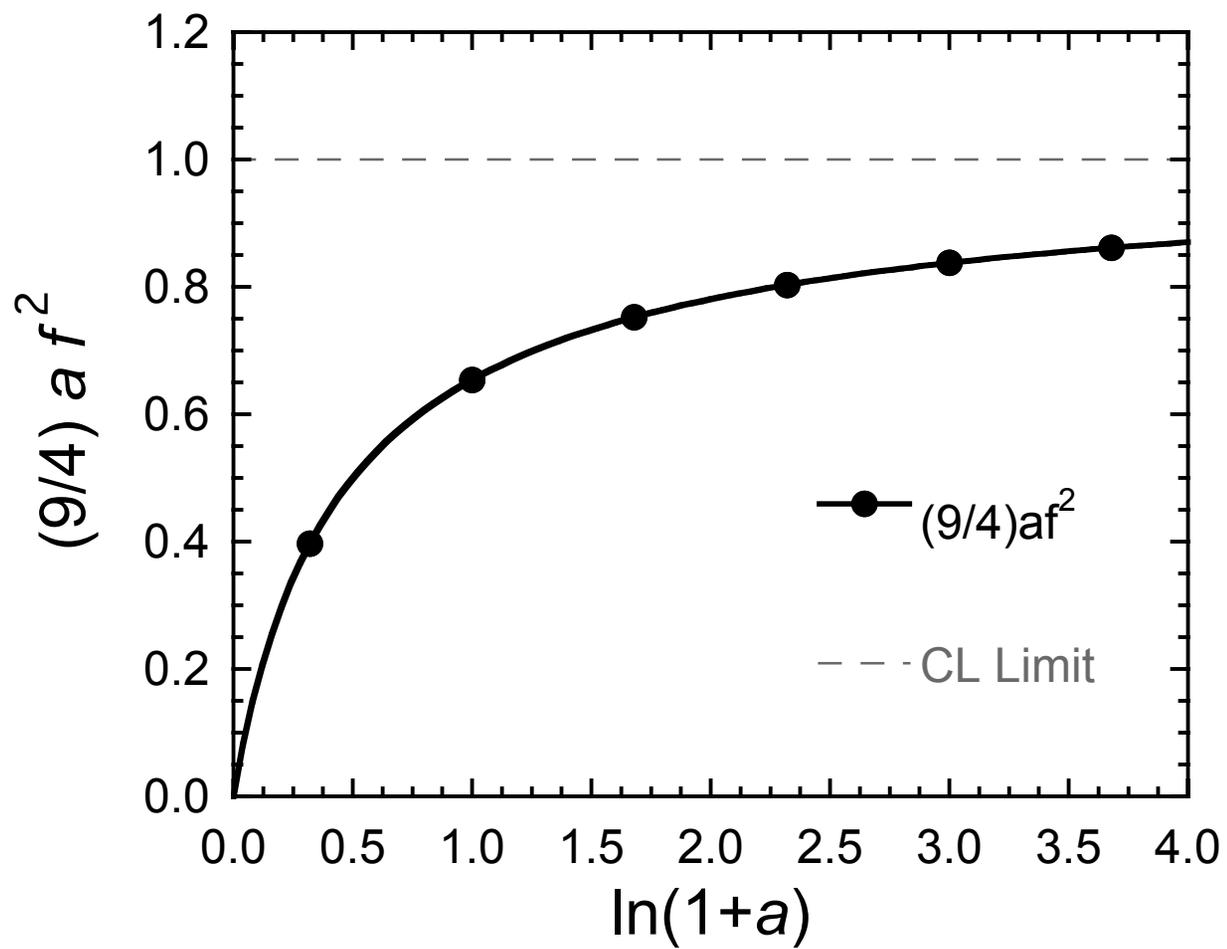

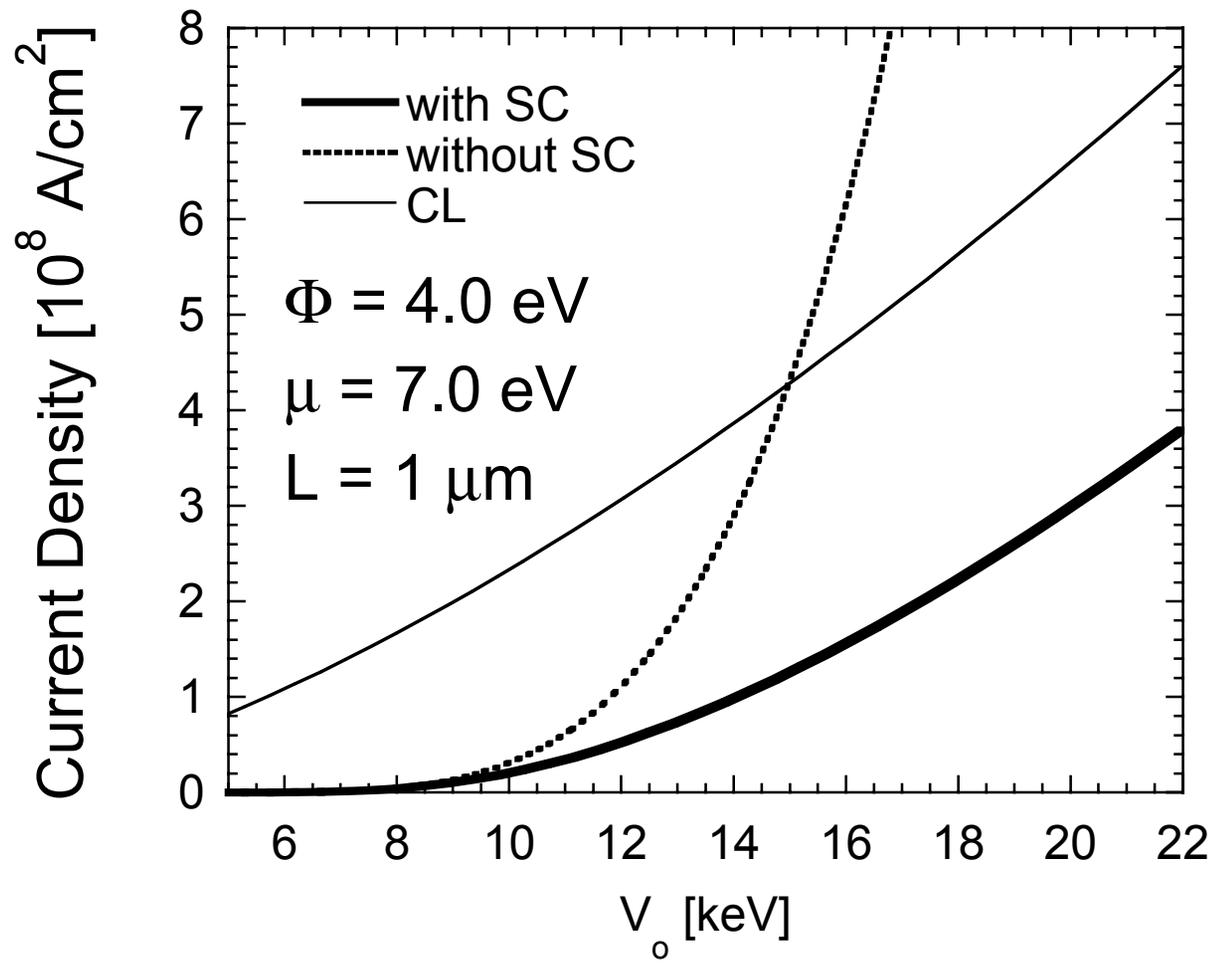

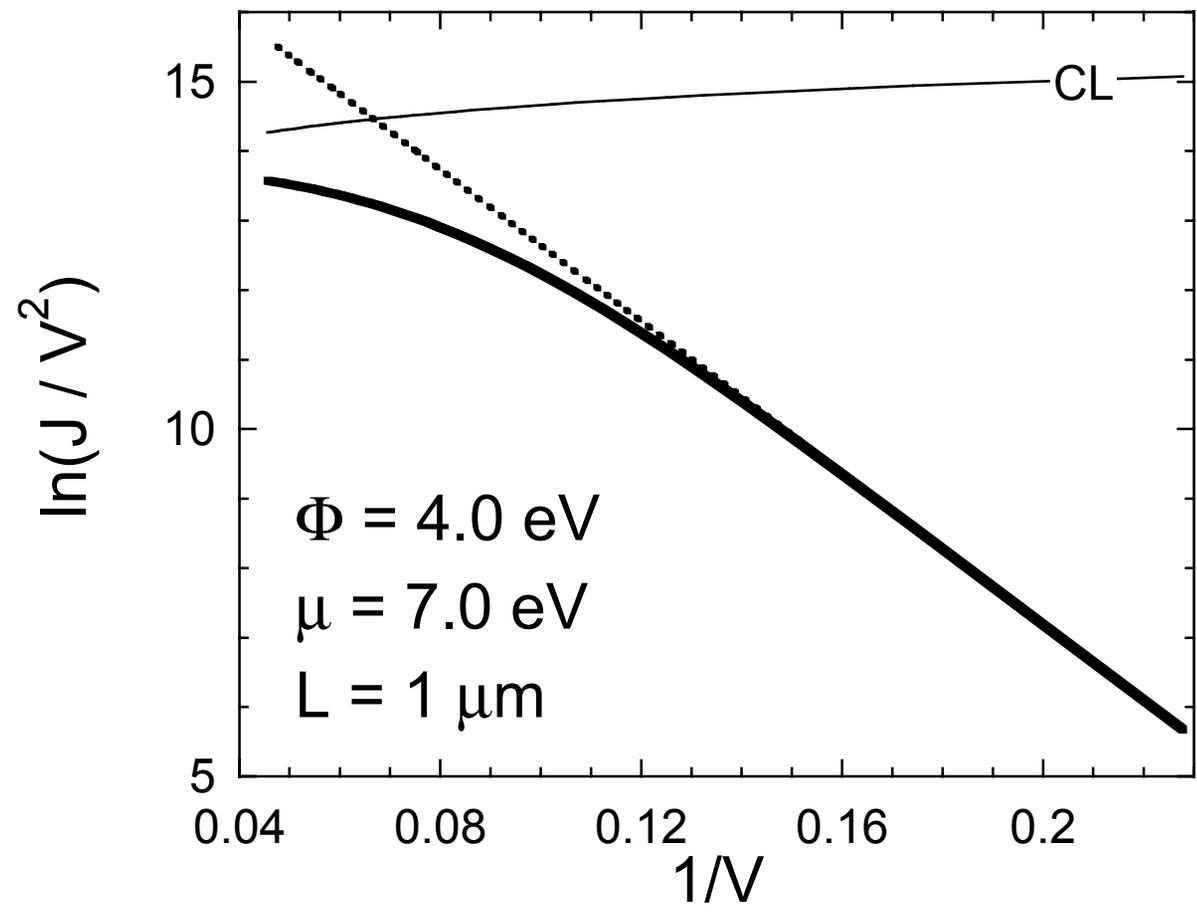

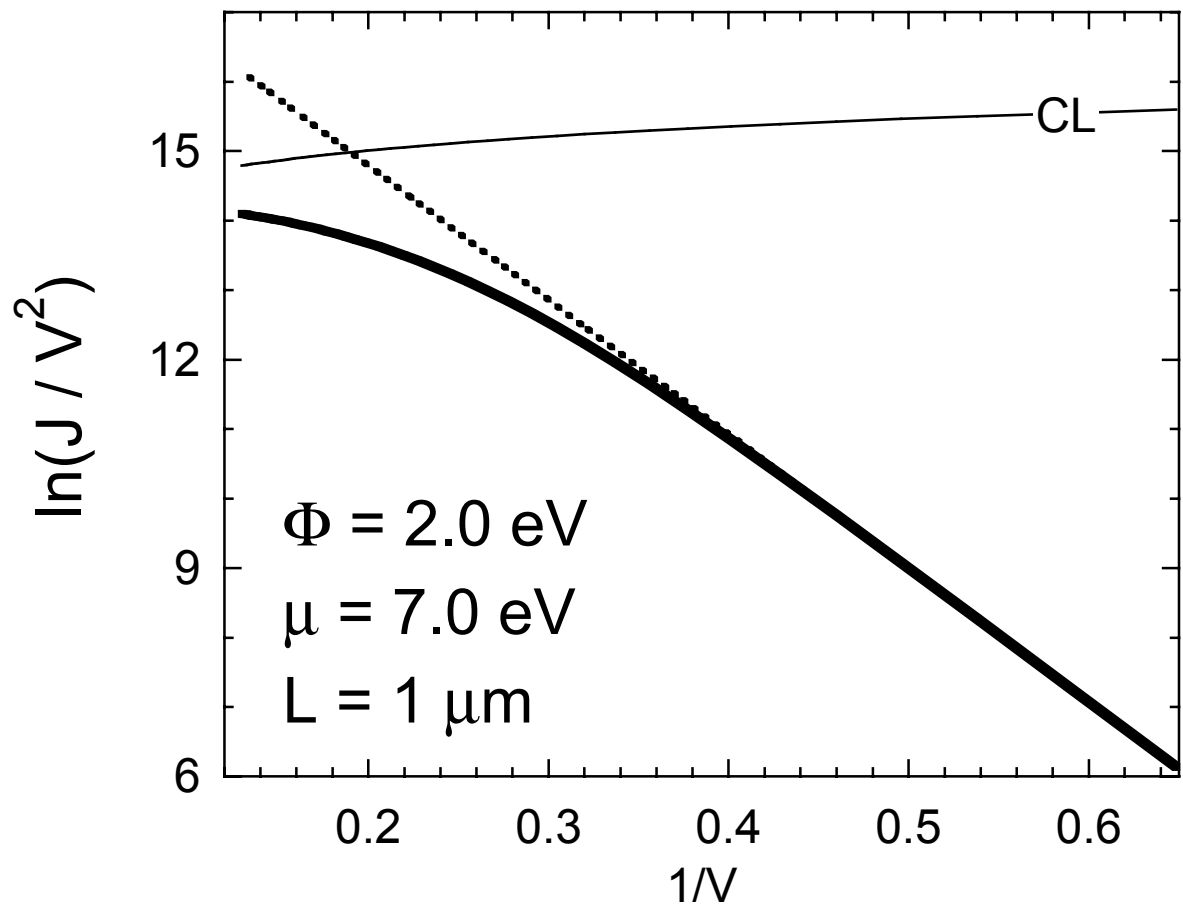

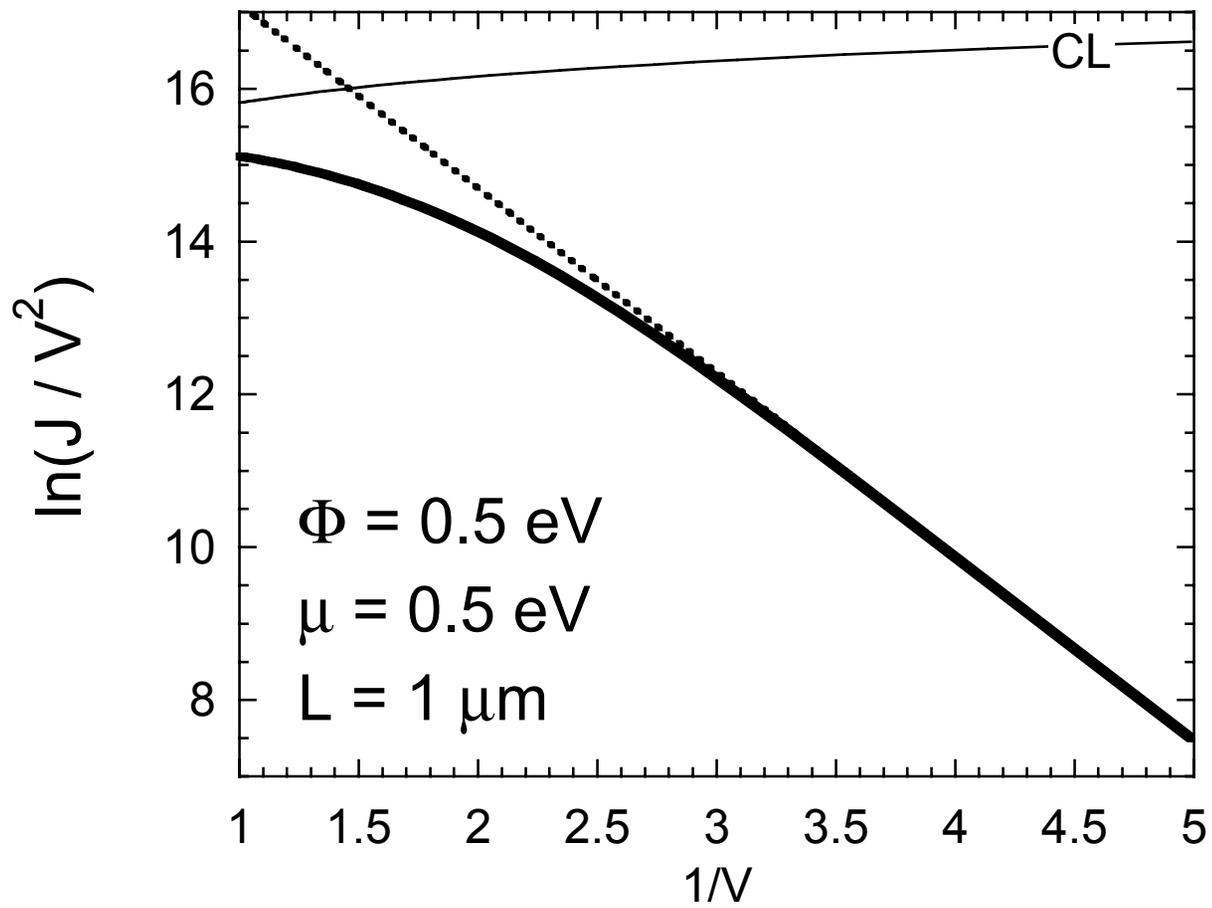

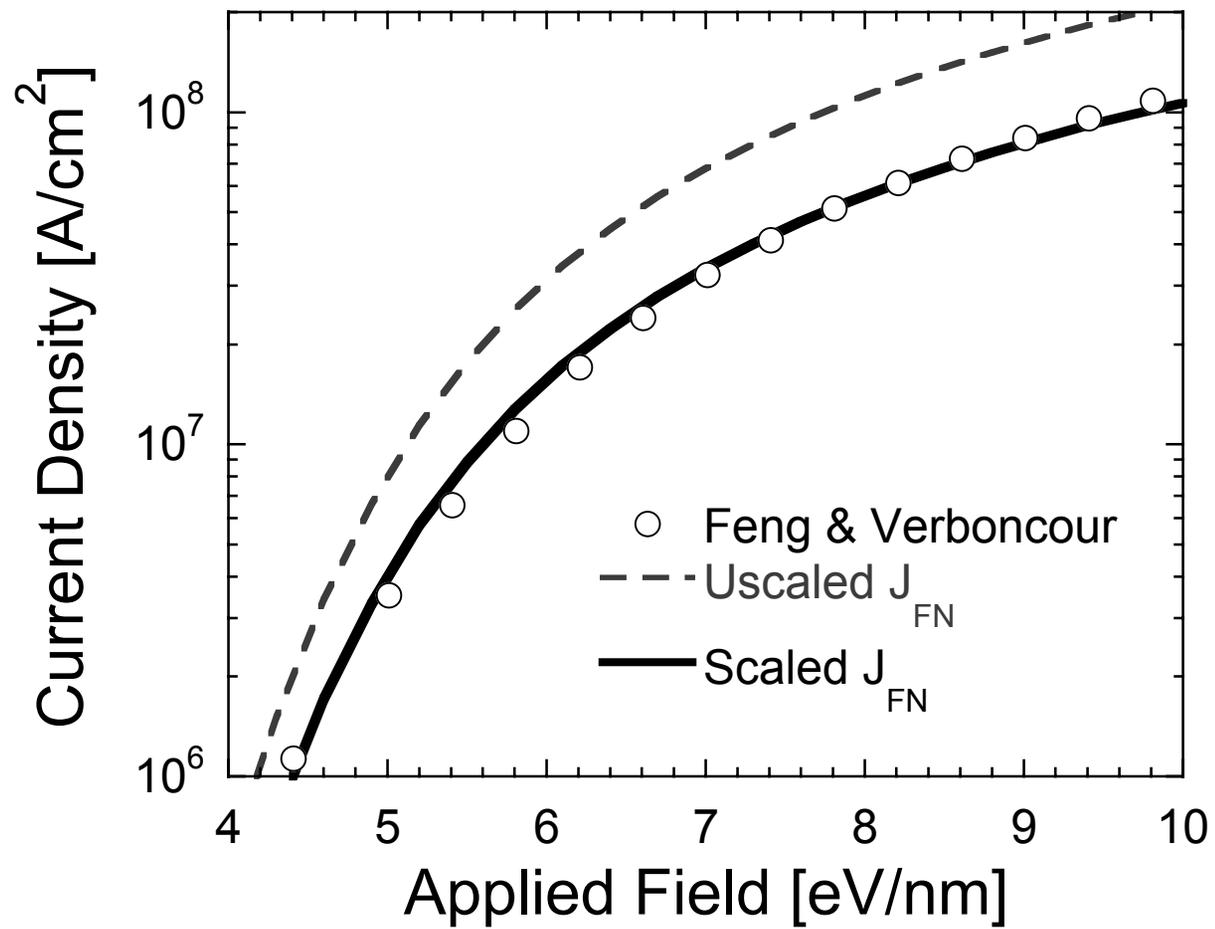

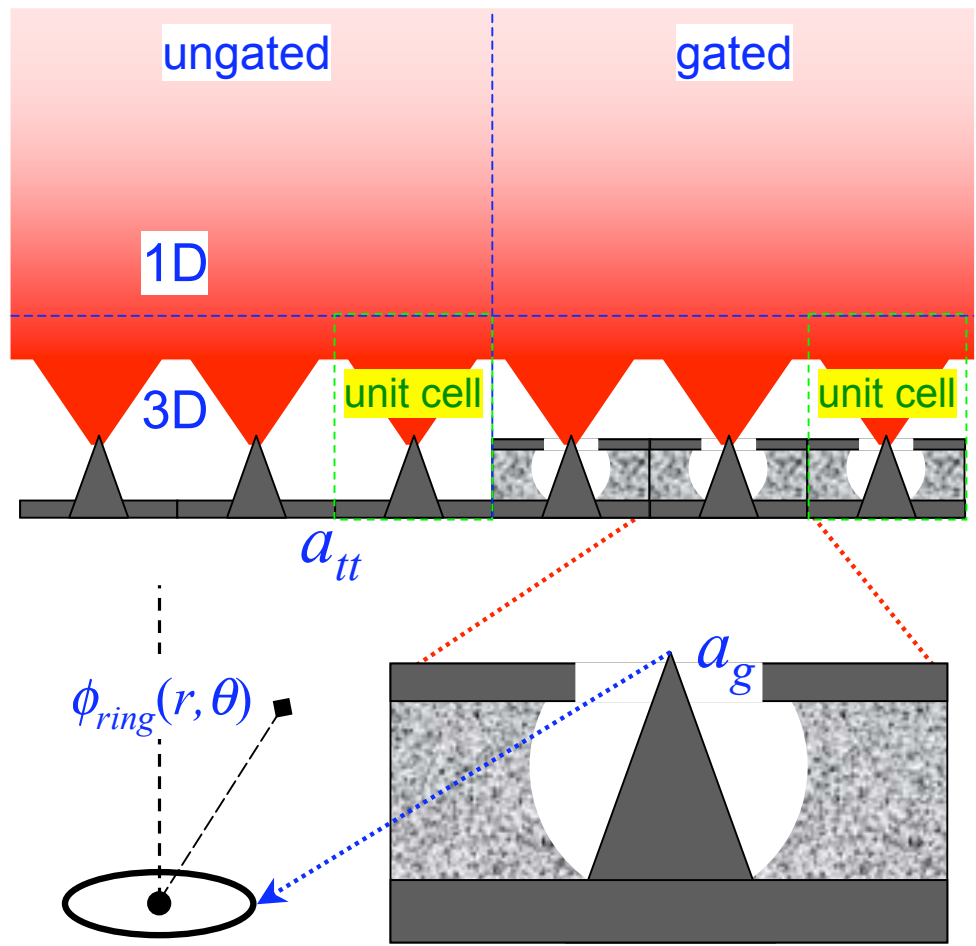

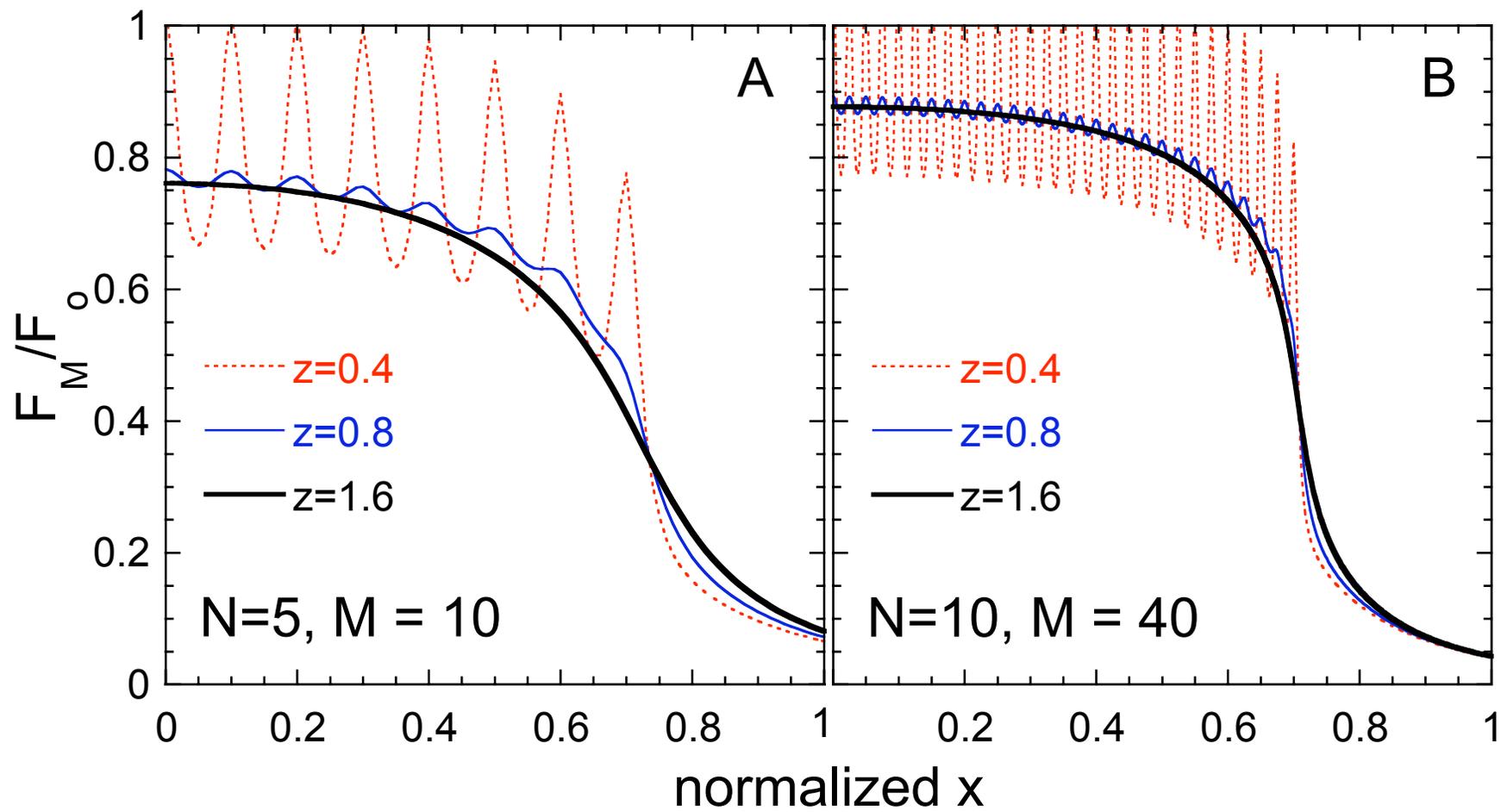

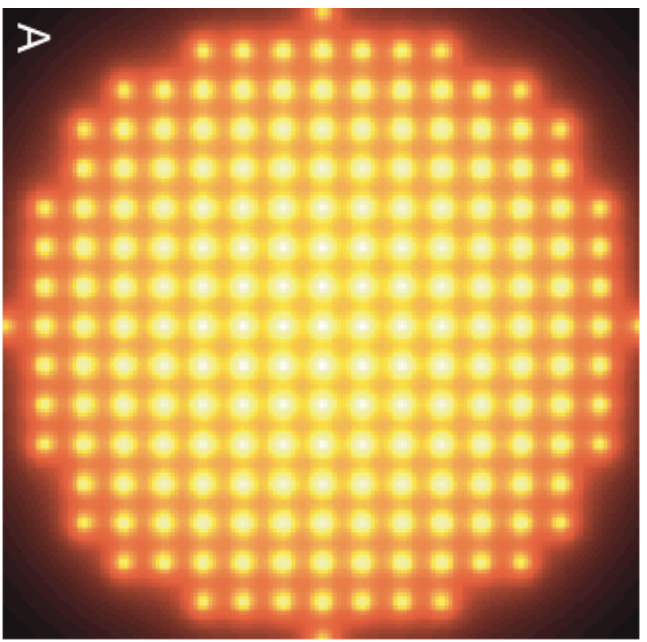
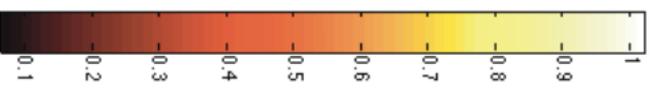
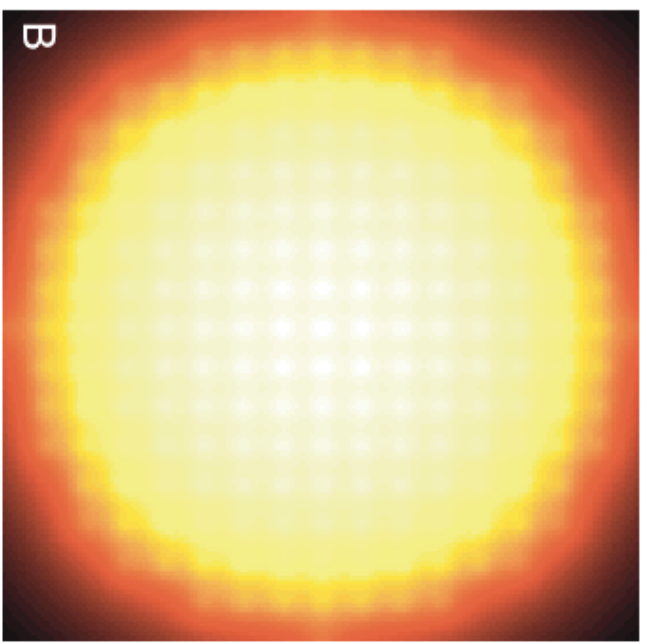
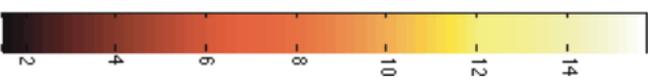

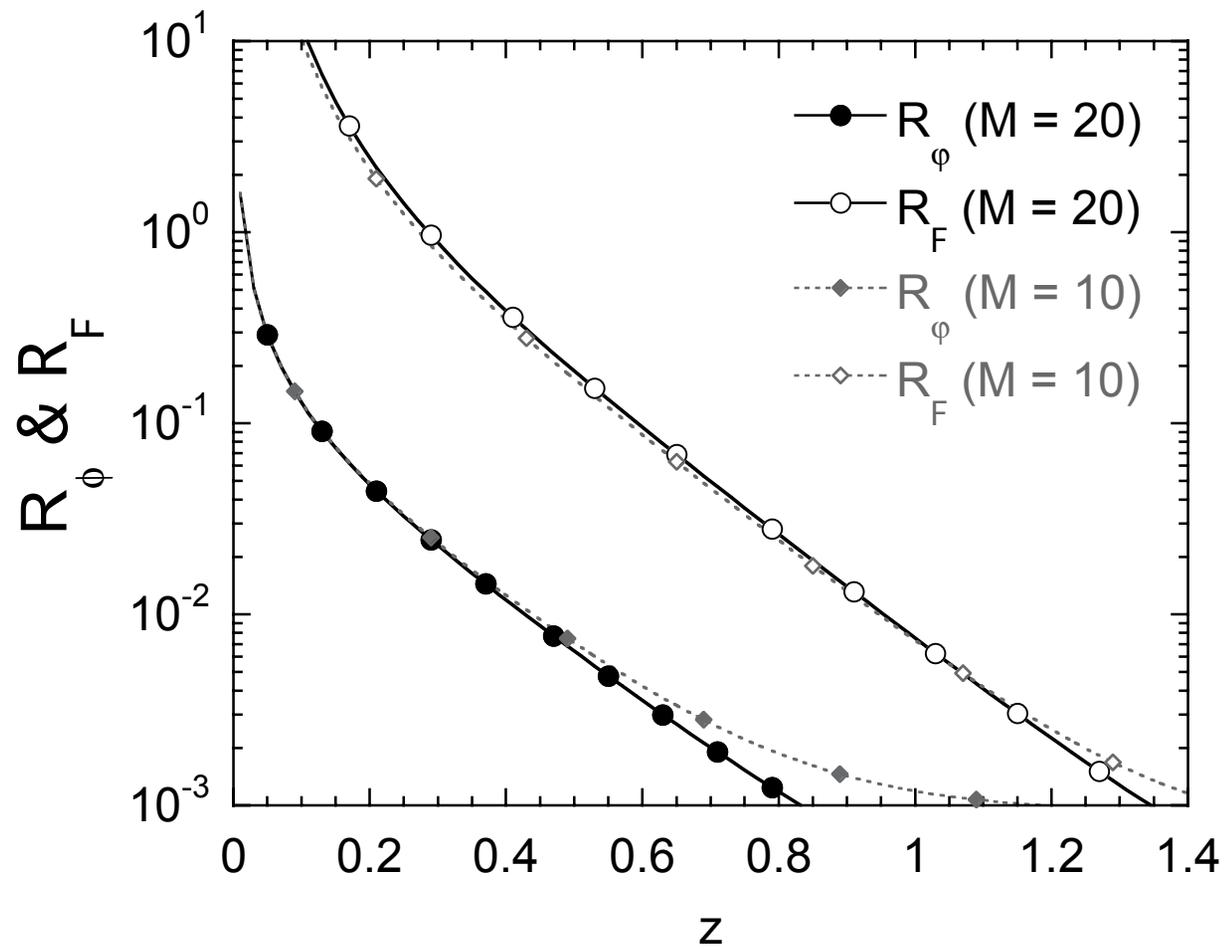

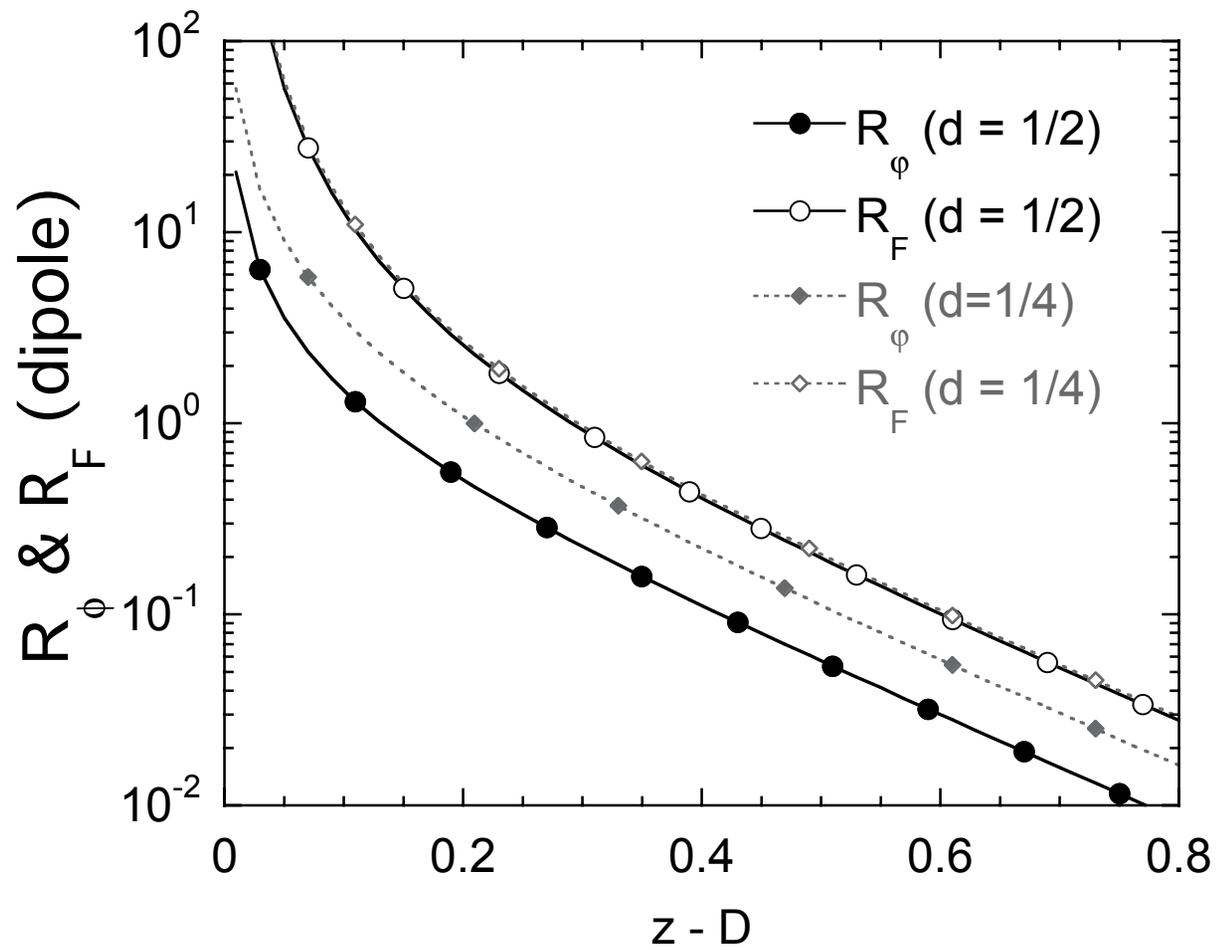

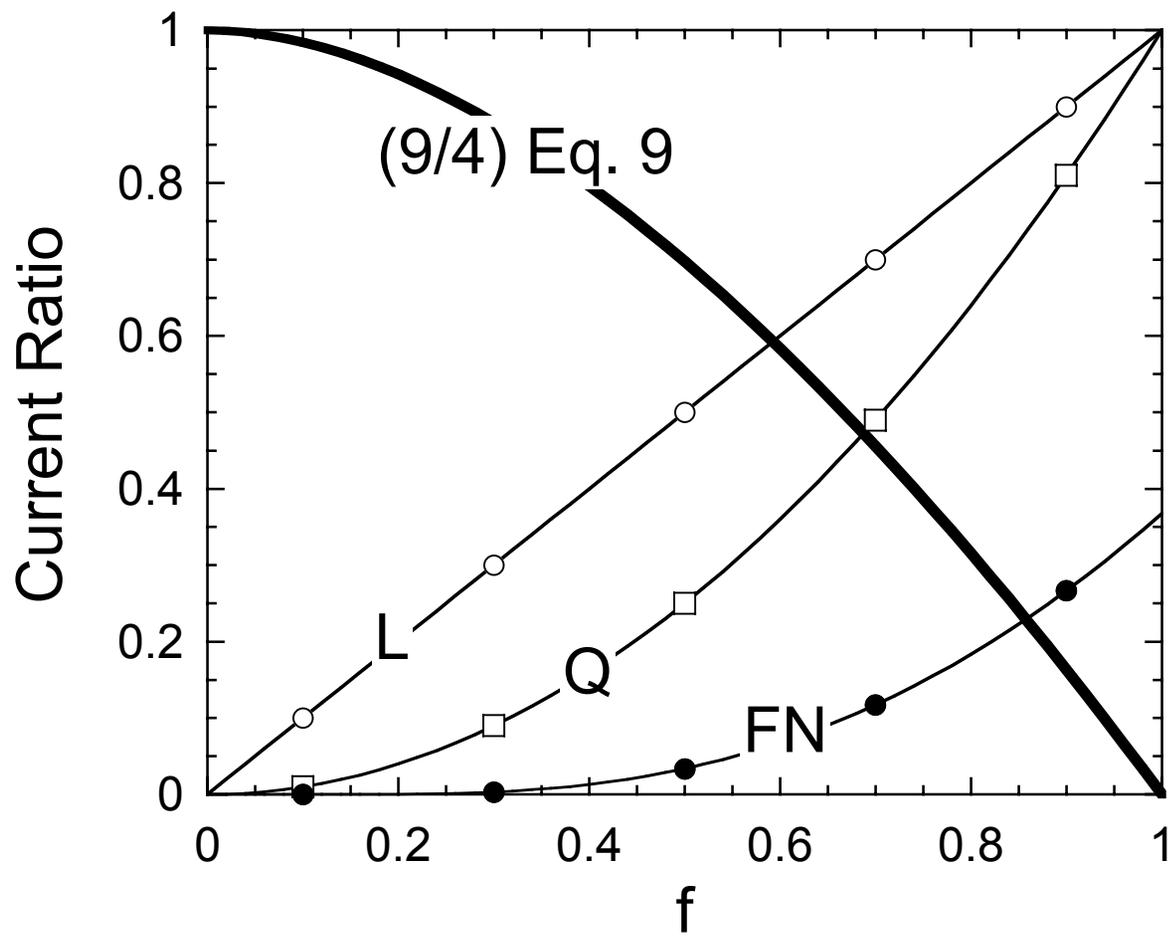